\documentclass[twocolumn]{aastex63}
\usepackage[varg]{txfonts}
\usepackage{graphicx}
\begin{document}

\title{No impact of core-scale magnetic field, turbulence, or velocity gradient on sizes of protostellar disks in Orion~A}

\author{Hsi-Wei Yen}
\affiliation{Academia Sinica Institute of Astronomy and Astrophysics, 11F of Astro-Math Bldg, 1, Sec. 4, Roosevelt Rd, Taipei 10617, Taiwan}

\author{Bo Zhao}
\affiliation{Department of Physics and Astronomy, McMaster University, Hamilton, Ontario, L8S 4M1, Canada}
\affiliation{Max-Planck-Institut f\"ur extraterrestrische Physik (MPE), Garching, Germany, 85748}

\author{Patrick M. Koch}
\affiliation{Academia Sinica Institute of Astronomy and Astrophysics, 11F of Astro-Math Bldg, 1, Sec. 4, Roosevelt Rd, Taipei 10617, Taiwan}

\author{Aashish Gupta}
\affiliation{Graduate Institute of Astronomy, National Central University, 300 Zhongda Road, Zhongli, Taoyuan 32001, Taiwan}
\affiliation{Academia Sinica Institute of Astronomy and Astrophysics, 11F of Astro-Math Bldg, 1, Sec. 4, Roosevelt Rd, Taipei 10617, Taiwan}

\correspondingauthor{Hsi-Wei Yen}
\email{hwyen@asiaa.sinica.edu.tw}

\begin{abstract}
We compared the sizes and fluxes of a sample of protostellar disks in Orion~A measured with the ALMA 0.87 mm continuum data from the VANDAM survey with the physical properties of their ambient environments on the core scale of 0.6 pc estimated with the GBT GAS NH$_3$ and JCMT SCUPOL polarimetric data. We did not find any significant dependence of the disk radii and continuum fluxes on a single parameter on the core scale, such as the non-thermal line width, magnetic field orientation and strength, or magnitude and orientation of the velocity gradient. Among these parameters, we only found a positive correlation between the magnitude of the velocity gradient and the non-thermal line width. 
Thus, the observed velocity gradients are more likely related to turbulent motion but not large-scale rotation.
Our results of no clear dependence of the disk radii on these parameters are more consistent with the expectation from non-ideal MHD simulations of disk formation in collapsing cores, where the disk size is self-regulated by magnetic braking and diffusion, compared to other simulations which only include turbulence and/or a magnetic field misaligned with the rotational axis. 
Therefore, our results could hint that the non-ideal MHD effects play a more important role in the disk formation. 
%, which self regulated the disk radii by magnetic braking and diffusion. 
Nevertheless, we cannot exclude the influences on the observed disk size distribution by dynamical interaction in a stellar cluster or amounts of angular momentum on the core scale, which cannot be probed with the current data.
\end{abstract}


\section{Introduction}
Protostellar disks around young protostars are potential sites of planet formation \citep{Manara18, Sheehan18, Cox20}.
Large samples of protostellar disks have been observed in thermal dust continuum emission \citep[e.g.,][]{Cox18, Cieza19, Maury19, Tobin20}.
Keplerian rotation of several protostellar disks around Class 0 and I protostars has also been detected and resolved \citep[e.g.,][]{Lommen08, Lee10, Murillo13, Ohashi14, Yen14, Lee18, Hsieh19}.
The sizes of these protostellar disks can differ by more than an order of magnitude \citep{Yen17,Maret20}.
The origins of these diverse disk sizes remain unclear. 
The difference in the disk sizes could be due to time evolution and/or suppression of disk growth in some protostellar sources but not in the others \citep{Li14, Tsukamoto20, Zhao20}. 

The growth of a protostellar disk is related to the transfer of mass and angular momentum from its parental dense core to the vicinity of its central protostar via collapsing material.
In hydrodynamics, where the angular momentum of collapsing material in dense cores is more or less conserved, 
the size of a protostellar disk is proportional to the rotation of its parental dense core and the mass which the disk has accreted \citep{Terebey84, Basu98}.
In ideal magnetohydrodynamics (MHD), 
the formation and growth of a protostellar disk can be severely suppressed because of efficient magnetic braking, 
when the rotational axis of the parental dense core is aligned with its magnetic field \citep{Allen03, Galli06, Mellon08}. 
Incorporating turbulence and a magnetic field misaligned with the rotational axis in dense cores can reduce the efficiency of magnetic braking and prompt formation of sizable protostellar disks in ideal MHD simulations \citep{Hennebelle06,Santos12,Joos12,Joos13,Li13,Li14b,Seifried13}. 
Non-ideal MHD effects with sufficient magnetic diffusivities, which are related to the cosmic-ray ionization rate and dust grain size distributions in dense cores \citep[e.g.,][]{Padovani14, Wurster16, Zhao16}, can also enable the formation of sizable protostellar disks \citep{Tsukamoto15a,Tsukamoto15b,Masson16,Zhao18,Wurster19,Hennebelle20}.
However, observationally it is still unknown which mechanism plays a more important role in the disk formation and evolution.

Several surveys with various single-dish telescopes and interferometers toward nearby star-forming regions have been conducted recently and provide rich data sets to study magnetic field, turbulence, and gas kinematics on multiple scales from dense cores ($\sim$0.1 pc), to protostellar envelopes ($\sim$1000 au), and to protostellar disks ($\sim$100 au) around young protostars \citep[e.g.,][]{Storm14,Friesen17,bistro,Stephens18,Stephens19,Cieza19,Tobin19}.
Observational comparisons of the sizes of protostellar disks with the properties of magnetic field, turbulence, and gas motions in dense cores could hint at the key mechanisms in the disk formation and evolution.
Therefore, in this work, we make use of publicly available survey data and compare the properties of a sample of protostellar disks in the Orion star-forming regions measured with the Atacama Large Millimeter/submillimeter Array (ALMA) with the physical conditions in their ambient environment observed with single-dish telescopes. 
We then discuss the dependences of the disk properties on the magnetic field, turbulence, and gas motions on the scale of dense cores to investigate the key mechanisms regulating the disk formation and evolution. 

\section{Sample}
The sample in the present paper is selected from the sample in the VLA/ALMA Nascent Disk and Multiplicity (VANDAM) Survey of Orion protostars \citep{Tobin19,Tobin20}.
The VANDAM survey of Orion is a survey toward all well-characterized protostars in the Orion~A and B molecular clouds with ALMA and toward a subsample of young protostars with the Very Large Array (VLA).
The total sample consists of 328 protostellar sources. 
The ALMA observations of the VANDAM survey are conducted at 0.87 mm and have a synthesized beam of 0\farcs11, corresponding to 39--49 au at the distances to Orion~A and B.
The Orion~A molecular clouds have also been mapped in the NH$_3$ emission with the Green Bank Ammonia Survey \citep[GAS;][]{Friesen17,Kirk17}, 
which is a survey to map the NH$_3$ emission at 23 GHz toward all the northern Gould Belt star forming regions with $A_V > 7$ with the Green Bank Telescope (GBT).
The GAS observations have a full-width-half-maximum beam size of 32$\arcsec$, corresponding to 11\,000--14\,000 au at the distances to Orion~A and B.

\begin{figure*}
\centering
\includegraphics[width=\textwidth]{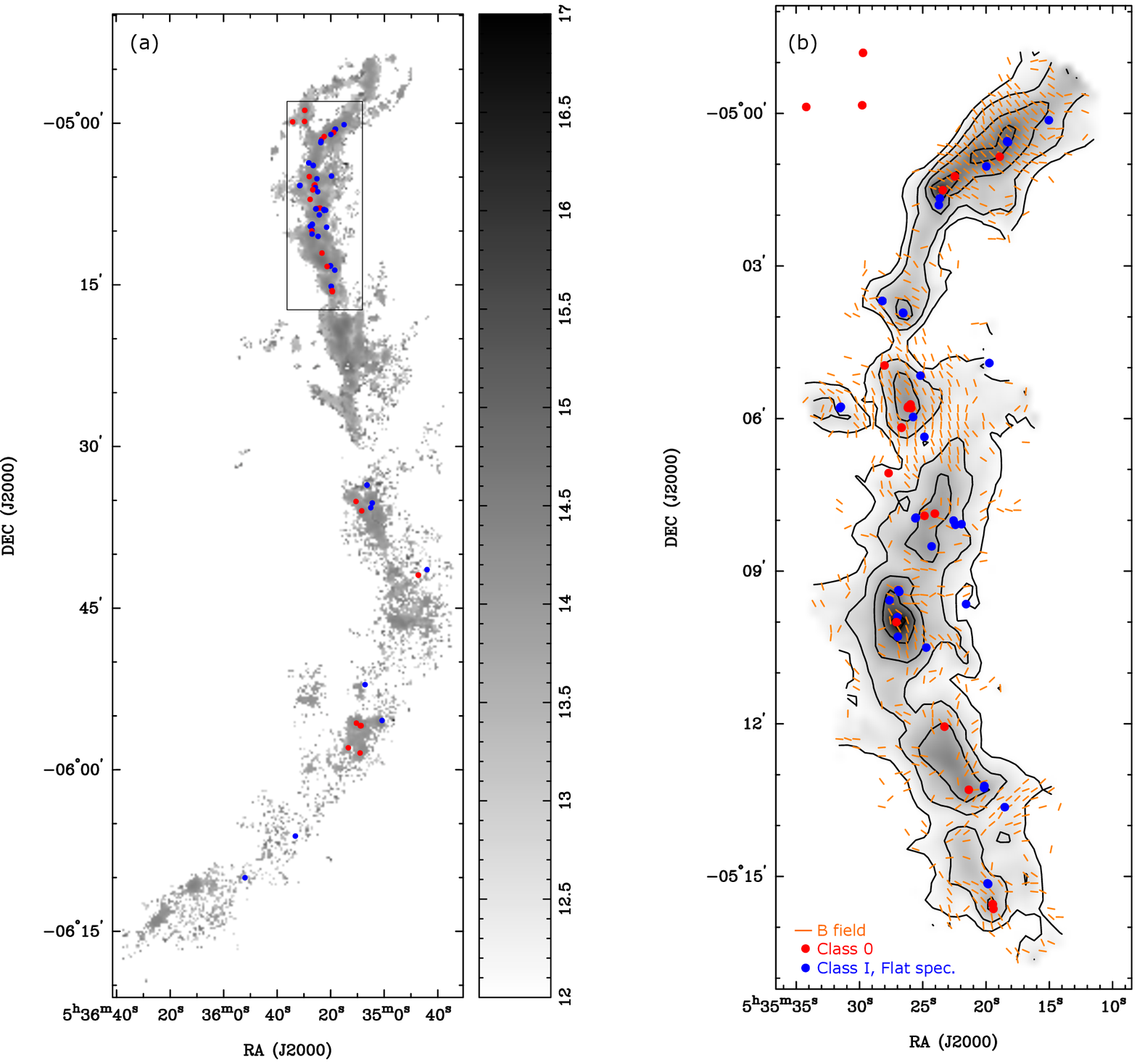}
\caption{(a) NH$_3$ column density map in a logarithmic scale of Orion~A obtained by the Green Bank Ammonia Survey \citep[GAS;][]{Friesen17,Kirk17}. Red and blue dots show the locations of our sample protostars selected from the VANDAM survey of Orion \citep{Tobin19, Tobin20}. A bolometric temperature of 70~K was adopted to separate Class 0 and more evolved protostars. Red dots are the Class 0 protostars, and blue dots are the Class I and flat-spectrum protostars. A rectangle denotes the mapping area of the JCMT SCUPOL survey \citep{Matthews09}. (b) 850 $\mu$m intensity map obtained with the JCMT SCUPOL survey. Orange segments show the magnetic field orientations inferred by rotating the polarization orientations by 90$\arcdeg$. Red and blue dots are the same as those in (a).}
\label{data_map}
\end{figure*}

From the VANDAM sample, we selected those protostellar sources which are detected in both the 0.87 mm continuum emission with ALMA and the NH$_3$ emission with GBT. 
This leads to a sample of 56 protostellar sources for the present study.
Among them, 25 are Class 0 protostars, 13 are Class I protostars, and 18 are flat-spectrum protostars \citep{Tobin20}. 
In addition, 26 out of the 56 sources are single protostars, and 30 are multiple systems, 
where single protostars are defined as having no other protostars within 10$^4$ au on the plane of the sky, the same as the definition in \citet{Tobin20}.
In total, there are 79 individual protostars in our sample. 
Figure~\ref{data_map}a presents the mapping area in Orion~A by GAS, 
and the locations of the individual protostars in our sample are labelled.
All the sample protostars are listed in Appendix~\ref{alldata}.

\section{Data and analysis}\label{analysis}
\subsection{ALMA 0.87 mm continuum}
We obtained the properties of the circumstellar disks around our sample protostars, including disk radii, orientations projected on the plane of the sky, and fluxes from Table~6 in \citet{Tobin20}.
In \citet{Tobin20}, an elliptical Gaussian was fitted to the 0.87 mm continuum emission in each detected source, 
and the deconvolved size and orientation and the total flux were reported. 
Following the definition in \citet{Tobin20}, 
we define the disk radius to be the 2$\sigma$ width of the fitted Gaussian function. 
For unresolved sources, 
the upper limits of their disk radii are adopted to be the 1$\sigma$ width of the synthesized beam, 
assuming that the observations are able to resolve disks with their diameters larger than the synthesized beam, which is $\sim$43 au at the distances to our sample protostars.
As discussed in \citet{Tobin20}, 
the contamination from protostellar envelopes is minimal in these measurements because large-scale envelopes are resolved out and the Gaussian fitting method also reduces the contribution by protostellar envelopes. 

\subsection{GBT NH$_3$ emission}
To measure gas kinematics and non-thermal motion on the core scale of 0.06 pc around our sample protostars, 
we retrieved the NH$_3$ property maps, including kinetic temperature, line width, and centroid line-of-sight velocity, of Orion~A from the publicly released datasets of GAS\footnote{\url{https://dataverse.harvard.edu/dataverse/GAS_Project}}.
For each protostar in our sample, 
we measured the mean kinematic temperature and line width over 3 by 3 pixels around the protostellar position. 
The pixel size of the NH$_3$ property maps is 10\farcs6 ($\sim$4200 au). 
The mean size of the dense cores in Orion~A identified in the JCMT 850 $\mu$m continuum map is 13\,000 au (or 0.06 pc) with a standard deviation of 6900 au \citep{Lane16,Kirk17}. 
Thus, our measurements averaged over 3 by 3 pixels can represent the mean properties on the typical core scale in Orion~A.
We did not perform a core identification using the NH$_3$ data, considering the limited resolution of the NH$_3$ data and the complex structures in Orion~A, as discussed in \citet{Lane16} and \citet{Kirk17}. 
We also found that a few protostars in our sample are not associated with the dense cores identified with the JCMT 850 $\mu$m continuum data by \citet{Lane16}.  
Nevertheless, we simply measured the properties over the core scale of 0.06 pc uniformly for all the sample protostars. 
Then the non-thermal line width ($\delta V_{\rm nt}$) on the core scale around each protostar was computed as 
\begin{equation}
\delta V_{\rm nt} = \sqrt{{\delta V_{\rm ob}}^2 - \frac{kT_{\rm k}}{m_{\rm NH_3}}},
\end{equation}
where $\delta V_{\rm ob}$ is the observed NH$_3$ line width, $k$ is the Boltzmann constant, and $m_{\rm NH_3}$ is the molecular weight of NH$_3$.
We also measured the magnitude and direction of the velocity gradient on the core scale around each protostar. 
We followed the method in \citet{Goodman93} and fitted a linear velocity gradient to 3 by 3 pixels around the protostellar position in the centroid line-of-sight velocity map.

\subsection{JCMT polarized 850 $\mu$m continuum}
The northern part of the Orion~A molecular clouds has been observed in the polarized thermal dust emission at 850 $\mu$m and 450 $\mu$m by the SCUPOL survey using the James Clerk Maxwell Telescope \citep[JCMT;][]{Matthews09}, as shown in Fig.~\ref{data_map}. 
To estimate the strengths and orientations of the magnetic field on the core scale around our sample protostars, 
we retrieved the Stokes {\it IQU} maps at 850 $\mu$m of OMC-2 and -3 obtained with the JCMT SCUPOL survey from the SCUBA Polarimeter Legacy Catalogue\footnote{\url{http://www.cadc-ccda.hia-iha.nrc-cnrc.gc.ca/community/scupollegacy/}}. 
These Stokes {\it IQU} maps were smoothed to have an angular resolution of 20$\arcsec$ \citep{Matthews09}.
We extracted polarization detections from the Stokes {\it IQU} maps with the detection criteria of the signal-to-noise ratios of Stokes {\it I} and polarized intensities both above three. 
Then we inferred the magnetic field orientations by rotating the polarization orientations by 90$\arcdeg$. 
The inferred magnetic field orientations are shown in Fig.~\ref{data_map}b. 

From the polarization detections, we computed mean Stokes {\it Q} and {\it U} intensities averaged over 3 by 3 pixels around each sample protostar, 
where the pixel size is 10$\arcsec$, 
and only the pixels with detections were included in the calculations. 
From the mean Stokes {\it Q} and {\it U} intensities, the mean magnetic field orientation on the core scale was derived for each protostar. 
In our sample, there are 56 protostars with nearby polarization detections within 15$\arcsec$ and thus with the measured mean magnetic field orientations on the core scale.

For the 34 protostars with more than three nearby polarization detections within 15$\arcsec$, 
we also measured its angle dispersion of the magnetic field orientations on the core scale of 0.06 pc. 
To remove contributions by large-scale magnetic field structures, 
we adopted the method similar to unsharp masking, as introduced in, for example, \citet{Pattle17} and \citet{WangJ19}.
For each polarization detection within 15$\arcsec$ from a sample protostar, 
we computed the mean Stokes {\it Q} and {\it U} intensities and thus the mean polarization orientation of 3 by 3 pixels centered at that detection, which is assumed to be the large-scale magnetic field orientation at the location of the detection, 
and only the pixels with detections were included in the calculations. 
Then this large-scale magnetic field orientation was subtracted from the orientation of that polarization detection. 
Finally, we calculated the standard deviation of the residual orientations of the polarizations detections within 15$\arcsec$ from the sample protostar, 
and obtained the angular dispersion of the magnetic field orientations on the core scale. 

We note that the calculated standard deviation can be biased due to the small number of detections included. 
We generated simulated samples with {\it N} numbers of data points randomly drawn from a Gaussian distribution, 
and calculated the standard deviations of the simulated samples. 
We found that the standard deviations tend to be underestimated by 13\%, 6\%, and 4\% when {\it N} is 4, 7, 10, 
and the uncertainties in the calculated standard deviations due to the small number of data points were estimated to be 39\%, 29\%, and 23\%, respectively.
We applied these corrections and uncertainties when we measured the angular dispersions of the magnetic field orientations on the core scale in our sample protostars based on the numbers of the polarization detections associated with them. 
As discussed in \citet{WangJ19}, we then subtracted the uncertainties in the polarization orientations due to the noise in the data ($\delta\theta_\sigma$) from the measured angular dispersion ($\delta\theta_{\rm ob}$) to estimate the intrinsic angular dispersions of the magnetic field orientations ($\delta\theta_{\rm B}$) as, 
\begin{equation}
\delta\theta_{\rm B} = \sqrt{{\delta\theta_{\rm ob}}^2-{\delta\theta_\sigma}^2}.
\end{equation}

Finally, we derived the magnetic field strengths on the core scale projected on the plane of the sky ($B_{\rm pos}$) in our sample protostars with the Davis--Chandrasekhar--Fermi method \citep{Davis51,Chandrasekhar53} as, 
\begin{equation}
B_{\rm pos} = \xi \sqrt{4 \pi \rho} \frac{\delta V_{\rm nt}}{\delta\theta_{\rm B}},  \label{DCF}
\end{equation}
where $\xi$ was adopted to be 0.5 to correct for inhomogeneous and complex magnetic field and density structures along the line of sight \citep{Ostriker01} and $\rho$ is the volume density.
$\delta V_{\rm nt}$ is from the GAS NH$_3$ data.
We estimated $\rho$ from the mean intensity of the 850 $\mu$m continuum emission on the core scale in the Stokes {\it I} maps obtained from the JCMT SCUPOL survey. 
For each sample protostar, we first measured the optical depth of the 850 $\mu$m continuum, assuming that the dust temperature is the same as the kinetic temperature measured with the GAS NH$_3$ data. 
Then we converted the optical depth to column density on the assumption of a dust mass opacity of 0.012 cm$^{2}$ g$^{-1}$, which includes a gas-to-dust mass ratio of 100.
These values are the same as those adopted in the studies of the dense cores in Orion~A \citep{Lane16,Kirk17}. 
The lengths of the clouds along the line of sight are unknown.
We assumed a constant length of 0.06 pc of all the sample sources and a 50\% uncertainty in the length, 
which are the mean and standard deviation of the diameters of the dense cores in Orion~A \citep{Lane16,Kirk17}, respectively.
With the estimated column density from the continuum emission and the assumed length along the line of sight, 
the volume density was derived for each sample protostar and adopted in Eq.~\ref{DCF}. 
The normalized mass-to-flux ratio ($\lambda$) was also computed for each sample protostar as, 
\begin{equation}
\lambda = 2\pi \sqrt{G}\frac{\Sigma}{B_{\rm pos}}, 
\end{equation}
where $G$ is the gravitational constant and $\Sigma$ is the column density \citep{Nakano78}.

\section{Results}
All the measurements obtained from the data and analysis described in Section~\ref{analysis} are listed in Appendix~\ref{alldata}.
In the following sections, we present the relations between the disk radii and other physical parameters on the core scale of 0.06 pc. 

\subsection{Potential biases due to multiplicity and evolution}

\begin{figure*}
\centering
\includegraphics[width=\textwidth]{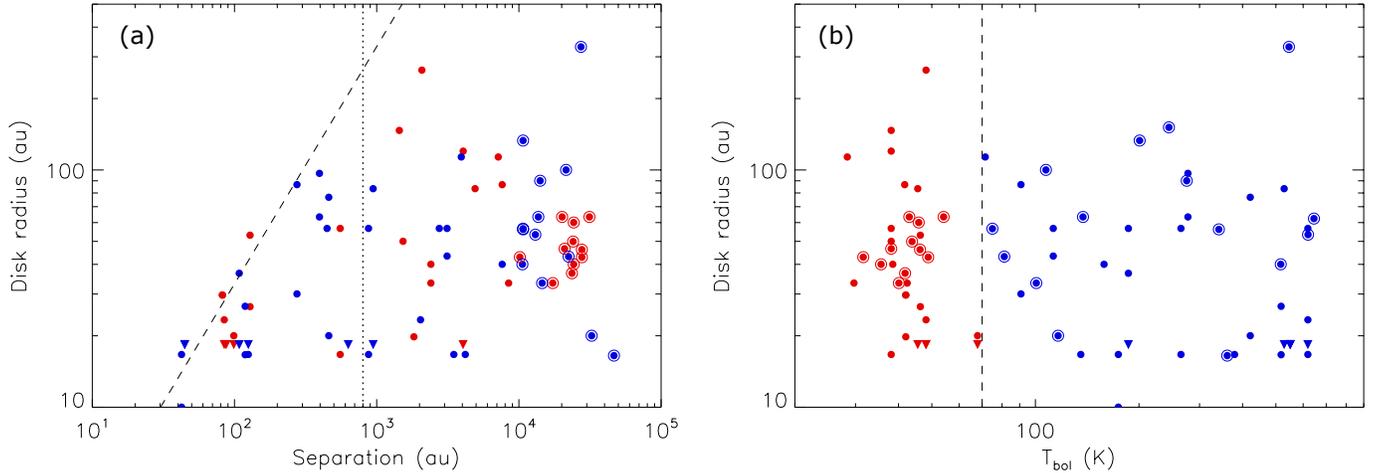}
\caption{Comparisons between the disk radii measured with the ALMA 0.87 mm continuum data and (a) the projected separations and (b) the bolometric temperatures of the protostars. If the disks are not resolved with the synthesized beam of 0\farcs11 (43 au), the upper limits of their radii are plotted as triangles. Red dots are the Class 0 protostars, and blue dots are Class I and flat-spectrum protostars. For a given protostar, the separation is defined as its distance to the protostar nearest to it. Single protostars are also plotted here with the separations calculated with the same definition, and they are labelled with open circles. In (a), a dashed line presents the expected truncation radius as a function of separation in a binary system with equal masses and a non-eccentric binary orbit, and a vertical dotted line denotes a separation of 800 au. In (b), a vertical dashed line denotes a bolometric temperature of 70 K, which is adopted to separate Class 0 and more evolved protostars.}
\label{sep}
\end{figure*} 

Our sample includes protostars in multiple systems. 
Circumstellar disks in multiple systems can be truncated due to dynamical interaction between companions \citep[e.g.,][]{Papaloizou77,Artymowicz94,Pichardo05}.
Figure~\ref{sep}a presents the disk radius as a function of projected separation on the plane of the sky.
For a given protostar, the separation is defined as its distance to the protostar nearest to it. 
Single protostars in our sample are also plotted for comparison following the same definition of the separation. 
The disks with smaller separations tend to have smaller radii. 
We calculated the expected truncation radius in a binary system with equal masses and a non-eccentric binary orbit following the equations in \citet{Manara19}.
The expected truncation radius as a function of separation in such binary systems is shown in Fig.~\ref{sep}a and approximately matches with the maximum disk radius observed at a given separation when the separation is smaller than 800 au. 
For the protostars with separations larger than 800 au, the expected truncation radii are more than a factor of two larger than their disk radii.
Thus, the disk radii around the protostars with separations smaller than 800 au are likely affected by the dynamical interaction in multiple systems, 
and the disks with the larger separations are not.
In the following discussions, we exclude the protostars with separations smaller than 800 au from our sample to avoid a bias due to binary interaction. 
The remaining sample consists of 50 individual protostars. 
We have confirmed that our discussions are not sensitive to this selection criterion based on the separation, and changing this selection criterion to a separation of 500 or 1000 au does not affect our results and conclusions.

The truncation radius can be smaller if the mass ratio is not unity or the binary orbit is eccentric in a binary system, which could result in small disks \citep{Artymowicz94,Pichardo05}.
This possibility cannot be ruled out in our sample. 
Nevertheless, we have compared the distributions of the disk radii of those singles and multiples in the subsample after excluding the protostars with separations smaller than 800 au. 
There is no statistically significant difference between the disk radius distributions of the singles and multiples. 
In addition, the ratios between the disk radii and the separations of the multiples in the subsample are typically less than 0.02 and at most 0.13. 
The theoretical calculations show that a low ratio of 0.02 requires a highly eccentric binary orbit with an eccentricity larger than 0.8 \citep{Manara19}.   
Highly eccentric binary systems are not common \citep[$<$30\%;][]{Duchene13,Tokovinin16,Tokovinin20}.
Thus, we expect that the disk radius distribution in our subsample after excluding protostars with separations smaller than 800 au is not biased by any possibly remaining binary interaction beyond 800 au.

We note that the ALMA 0.87 mm continuum emission only traces dusty disks. 
Sizes of dusty disks can be different from those of gaseous disks due to different opacities between lines and continuum, dust drift, and grain growth \citep[e.g.,][]{Birnstiel10, Facchini17}.
Observationally, sizes of Class II disks are indeed often found to be larger in molecular lines than in the millimeter continuum \citep[e.g.,][]{Ansdell18,Yen18,Sanchis21}.
In the full VANDAM sample of Orion, 
\citet{Tobin20} found that the median disk radius decreases by 40\% (from 50 au to 30 au) from Class 0 to flat-spectrum protostars, 
and the disk radii observed at 7 mm are typically smaller than those observed at 0.87 mm.
These trends could suggest dust evolution in disks. 

Figure~\ref{sep}b presents the disk radii as a function of bolometric temperature ($T_{\rm bol}$) in our sample. 
$T_{\rm bol}$ can be an evolutionary indicator \citep{Ladd93,Chen95}. 
There is no clear sign of decreasing disk radii with time evolution, possibly due to the smaller sample size, as compared to the full VANDAM sample. 
The median disk radii of the Class 0, Class I, and flat-spectrum in our subsample are 50 au, 57 au, and 40 au, respectively. 
On the contrary, the disk radii of protostars at similar evolutionary stages can differ by an order of magnitude. 
Thus, dust evolution is unlikely a dominant mechanism in determining the disk size distribution observed in the millimeter continuum. 
In addition, in a few embedded Class 0 and I protostars, where the disk radii were measured with gas kinematics, 
the disk radii observed in the millimeter continuum are comparable to the dynamically measured disk radii within a factor of two \citep[e.g.,][]{Ohashi14,Yen14,Aso15,Aso17,Sai20}.
This difference within a factor of two has also been found in the synthetic observations of the numerical simulations of a Keplerian disk formed in a magnetized collapsing core \citep{Aso20}.
Therefore, we expect that other mechanisms, such as initial angular momentum distributions, magnetic field, and/or turbulence in parental dense cores \citep[e.g.,][]{Terebey84,Basu98,Li14,Hennebelle16,Wurster20}, are more important in determining the disk radius distribution in our sample.

\subsection{Dependence on turbulence and magnetic field}

\begin{figure*}
\centering
\includegraphics[width=\textwidth]{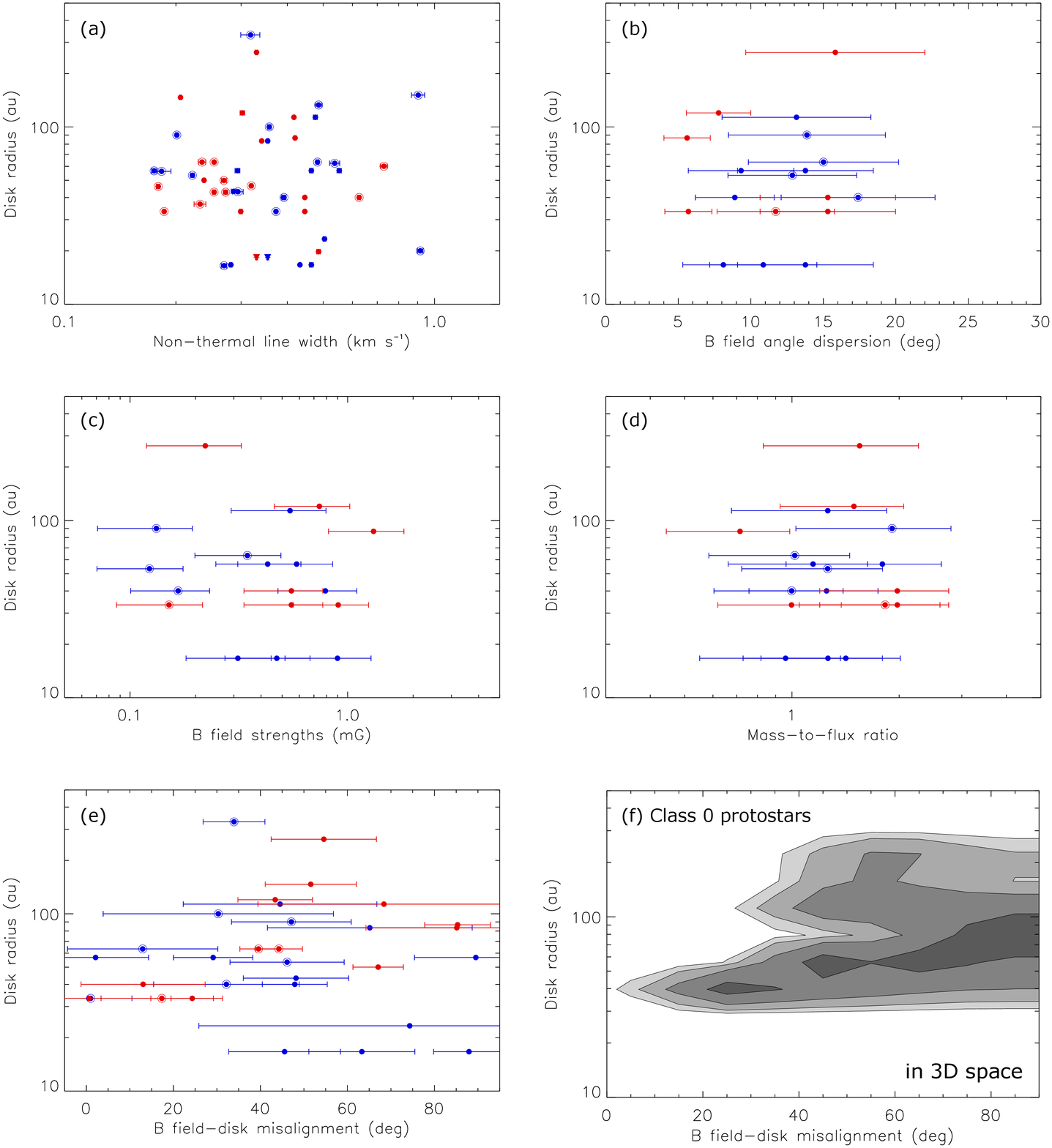}
\caption{Comparisons between the disk radii measured with the ALMA 0.87 mm continuum data and (a) the non-thermal line widths measured with the GBT NH$_3$ data, (b) the angular dispersions of the magnetic field orientations measured with the JCMT SCUPOL data, (c) the magnetic field strengths estimated with the Davis--Chandrasekhar--Fermi method, (d) the normalized mass-to-flux ratios, and (e) the misalignments between the rotational axes of the disks and the magnetic fields projected on the plane of the sky, on the core scale of 0.06 pc. Error bars present the 1$\sigma$ uncertainties. If the disks are not resolved with the synthesized beam of 0\farcs11 (43 au), the upper limits of their radii are plotted as triangles. Red dots are the Class 0 protostars, and blue dots are Class I and flat-spectrum protostars. Single protostars are labelled with open circles. (f) Inferred probability distribution of the 3D misalignments between the rotational axes of the disks and the magnetic fields for the Class 0 protostars in (e). The distribution was sampled with bin sizes of the disk radii in a logarithm scale from 1.45 in steps of 0.075 in units of au and the 3D misalignment angles in a linear scale from 0$\arcdeg$ in steps of 10$\arcdeg$. The probabilities to have a data point within the innermost, second, third, and outermost contours are 35\%, 81\%, 96\%, and 98\%, respectively, where the probability is computed as the number of the simulated data points enclosed by a given contour divided by the total number of all the data points.
%The contour levels are 0.5\%, 1\%, 2\% and 4\%, where the probability is computed as the number of the simulated data points in one bin divided by the total number of all the data points.
}
\label{tur_mag}
\end{figure*} 

Figure~\ref{tur_mag}a compares the disk radii with the non-thermal line widths, which are expected to trace the levels of the turbulence on the core scale. 
There is no correlation between the turbulence and the disk radii, neither in the entire subsample or only for the young protostars with $T_{\rm bol}<70$ K in the subsample (red data points).
Figure~\ref{tur_mag}b compares the disk radii with the angular dispersions of the magnetic field orientations on the core scale. 
The angular dispersion can be a proxy of the ratio between the energy densities of the magnetic field and the turbulence in the gas \citep{Davis51,Chandrasekhar53}. 
There is also no correlation between the disk radii and the angular dispersions of the magnetic field orientations in our sample. 
Similarly, the comparisons of the disk radii with the magnetic field strengths and the mass-to-flux ratios on the core scale also show that there is no dependences of the disk radii on these two parameters in our sample (Fig.~\ref{tur_mag}c and d). 
We computed the linear Pearson correlation\footnote{The linear Pearson correlation was computed with the IDL function CORRELATE, \url{https://www.l3harrisgeospatial.com/docs/correlate.html} \citep{Neter87}.} coefficients between the disk radii and these parameters of the turbulence and the magnetic fields. 
The absolute values of the correlation coefficients are mostly less than 0.1 and at most 0.17. 
We also computed the Spearman's rank correlation\footnote{The Spearman's rank was computed with the IDL function R\_CORRELATE, \url{https://www.l3harrisgeospatial.com/docs/r_correlate.html} \citep{William92}.}, which is more sensitive to a non-linear correlation. 
No significant correlations were found among these parameters, and the confidence levels of having correlations are all lower than 50\%, no matter we consider the entire subsample or only the young protostars with $T_{\rm bol}<70$ K in the subsample.

We also compared the disk radii with the misalignments between the rotational axes of the disks and the magnetic field orientations on the core scale projected on the plane of the sky. 
The rotational axes of our sample disks are defined as the axes perpendicular to the disk major axes measured with the ALMA observations, 
and the misalignments of 0$\arcdeg$ and 90$\arcdeg$ mean the magnetic fields are parallel and perpendicular to the rotational axes, respectively.
The disk radii in our entire subsample do not show any dependence on the misalignments. 
The linear Pearson correlation coefficient was computed to be 0.06, 
and the calculation of the Spearman's rank correlation also suggests that there is no significant correlation. 
We note that the misalignments presented in Fig.~\ref{tur_mag}e are angles projected on the plane of the sky. 
The actual misalignment angle in three-dimensional (3D) space has a higher probability to be the same as or larger than the projected angle \citep{Galametz20}.
Thus, some of the data points in Fig.~\ref{tur_mag}e may shift toward the right-hand side if we correct for the projection effects. 
Nevertheless, since there are several disks with radii smaller than 30 au and projected misalignment angles larger than 45$\arcdeg$, 
we do not expect that there would be an emerging trend in this subsample after we consider the projection effects.

When we only considered the young protostars with $T_{\rm bol} < 70$~K, 
we found a possible correlation between the disk radii and the misalignments between the magnetic fields and the rotational axes. 
The Pearson correlation coefficient was computed to be 0.4, 
and the calculation of the Spearman's rank correlation shows a correlation coefficient of 0.6 at a confidence level of 98\%. 
To test the significance of this correlation considering the projection effects, 
we followed calculations in \citet{Galametz20} to compute the probability distribution of the 3D misalignment angles of these 12 Class 0 protostars.
For each data point of the Class 0 protostars, we first inferred its 3D misalignment angle assuming that the magnetic field and the rotational axis are randomly oriented in 3D space, 
and the measured disk radius was not affected by projection.
We repeated this process for 5000 times, so we had 60\,000 simulated data points.
Then we plotted the probability distribution of the 3D misalignment angles versus the disk radii of these simulated data points (Fig.~\ref{tur_mag}f).
For each iteration having 12 simulated data points, we also computed the linear Pearson correlation coefficient. 
From the distribution of all the computed coefficients, we estimated the linear Pearson correlation coefficient between the misalignment angles and the disk radii to be 0.2$\pm$0.2 after considering the projection effects.
Although the lack of large disks with small misalignment angles may hint at a correlation between the misalignment and the disk radius, as shown in Fig.~\ref{tur_mag}e and \ref{tur_mag}f, 
with the current sample size this correlation is not significant ($\sim$1$\sigma$) after considering the projection effects.
Especially, the probability distribution is almost flat when the 3D misalignment angles are larger than 30$\arcdeg$ (Fig.~\ref{tur_mag}f).

%However, we found that the presence of this correlation is driven by the data point with an almost zero misalignment angle and a disk radius of 30 au. 
%If this data point is removed from the calculations, there is no significant correlation. 
%The Pearson correlation coefficient becomes 0.23, 
%and the confidence level computed with the Spearman's rank correlation drops below 90\%.
%Considering the projection effect, 
%the probability distribution of the actual misalignment angle in the 3D space of this data point is close to uniform from 0$\arcdeg$ to 90$\arcdeg$. \citep{Galametz20}. 
%Thus, this tentative correlation between the disk radii and the misalignments in the young protostars is not significant. 

\subsection{Dependence on velocity gradient}

\begin{figure*}
\centering
\includegraphics[width=\textwidth]{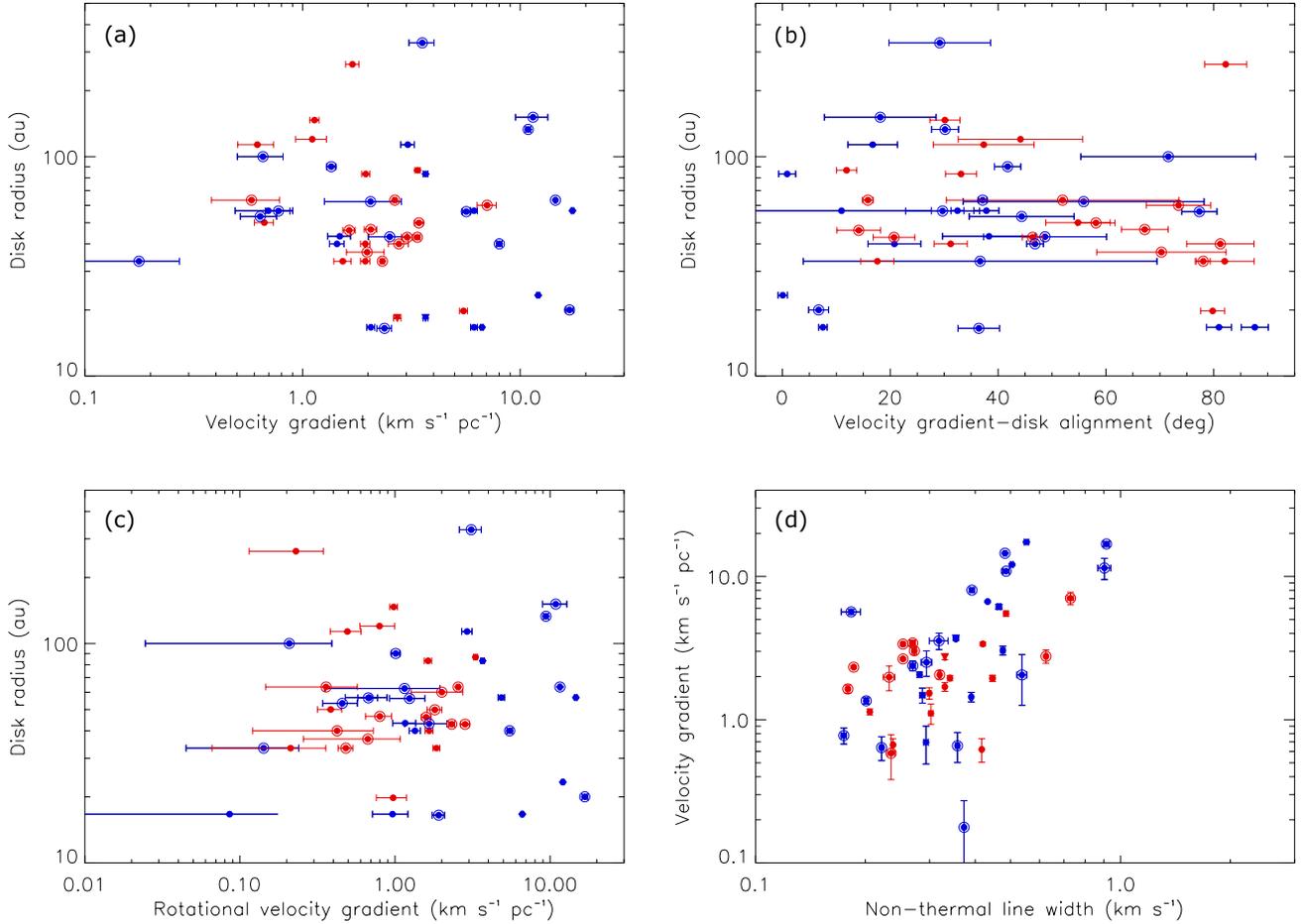}
\caption{Comparisons between the disk radii measured with the ALMA 0.87 mm continuum data and (a) the magnitudes of the velocity gradients measured with the GBT NH$_3$ data, (b) the alignments between the disk major axes and the velocity gradients, and (c) the magnitudes of the velocity gradients projected onto the disk major axes as an assessment of possible rotational motion, on the core scale of 0.06 pc. Panel (d) compares the magnitudes of the velocity gradients with the non-thermal line widths on the core scale measured with the GBT NH$_3$ data. Error bars present the 1$\sigma$ uncertainties. If the disks are not resolved with the synthesized beam of 0\farcs11 (43 au), the upper limits of their radii are plotted as triangles.  Red dots are the Class 0 protostars, and blue dots are Class I and flat-spectrum protostars. Single protostars are labelled with open circles.}
\label{vg}
\end{figure*} 

Figure~\ref{vg}a compares the disk radii with the magnitudes of the velocity gradients on the core scale.
%Conventionally, velocity gradients in dense cores are considered to trace the rotation of dense cores \citep[e.g.,][]{Goodman93}, 
%and faster rotations in dense cores can result in larger disks around central protostars \citep{Terebey84, Basu98}. 
Our results show that there is no dependence of the disk radii on the magnitude of the velocity gradients on the core scale. 
We also compared the disk radii with the alignments between the disk major axes and the direction of the velocity gradients. 
Rotational motion can induce a velocity gradient along the disk major axis, which is perpendicular to the rotational axis. 
On the assumption that the velocity gradients trace systematic motions on the core scale, 
rotational motion could be more dominant on the core scale around the disks with smaller alignment angles. %between the disk major axes and the velocity gradients.
Nevertheless, there is also no correlation seen in Fig.~\ref{vg}b.
To assess the possible amount of rotation on the core scale, 
we projected the velocity gradients onto the directions along the disk major axes (Fig.~\ref{vg}c), 
and we also did not find any correlation between the disk radii and the magnitudes of the velocity gradients along the disk major axes. 
The linear Pearson correlation coefficients between the disk radii and these parameters related to the velocity gradients on the core scales were computed, 
and the absolute values were all below 0.1.

In addition to correlations with the disk radii, 
we also investigated correlations between the parameters of the turbulence, magnetic fields, and velocity gradients on the core scale. 
We found that the magnitudes of the velocity gradients are correlated with the non-thermal line widths (Fig.~\ref{vg}d) with a Pearson correlation coefficients of 0.68. 
The calculation of the Spearman's rank correlation also suggests a correlation coefficient of 0.56 at a confidence level higher than 99\%.
We have also compared the fluxes of the sample disks with the parameters of the turbulence, magnetic fields, and velocity gradients on the core scale.
Similar to the results of the disk radii, no significant correlation was found between the disk fluxes and these parameters. 

\section{Discussion}
\subsection{Effects of turbulence and magnetic field}\label{discuss1}
The magnetic field in protostellar sources is theoretically expected to suppress the formation and growth of circumstellar disks around protostars via magnetic braking \citep[e.g.,][]{Tomisaka00,Allen03,Galli06,Mellon08}.
Ideal MHD simulations demonstrate that the turbulence and the misalignment between the magnetic field and the rotational axis in dense cores can reduce the efficiency of magnetic braking and prompt formation of sizable Keplerian disks around protostars \citep{Hennebelle06,Santos12,Joos12,Joos13,Li13,Li14b,Seifried13,Gray18}. 
Our results show that the disk radii in Orion~A do not correlate with the non-thermal line widths, the angular dispersions of the magnetic field orientations, and the misalignment between the magnetic fields and the rotational axes on the core scale (Fig.~\ref{tur_mag}). 
Different from the ideal MHD simulations, 
the non-ideal MHD simulations suggest that the turbulence and the misalignment do not significantly affect properties of disks if the non-ideal MHD effects are present \citep{Masson16,Lam19,Wurster19a,Wurster20}. 
In these non-ideal MHD simulations, the radii of formed disks are all comparable regardless of the initial turbulence, misalignment, and magnetic field strengths in dense cores, when the initial magnetic field is relatively strong with a mass-to-flux ratio of a few \citep{Hennebelle16,Hennebelle20,Wurster20}.
This is because the disk radii are self-regulated by magnetic braking and diffusion of the magnetic field due to the non-ideal MHD effects, 
and faster rotation results in more toroidal magnetic fields and thus strong magnetic braking \citep{Hennebelle16,Hennebelle20,Zhao20}. 
The mass-to-flux ratios on the core scale in our sample protostars are estimated to be on the order of unity (Fig.~\ref{tur_mag}d), which falls into the regime of the strong magnetic fields in these non-ideal MHD simulations. 
Thus, the absence of a clear dependence of the disk radii on the initial turbulence, misalignment, and magnetic field strengths on the core scale in our results  is consistent with the expectation from these non-ideal MHD simulations, and could support that the non-ideal MHD effects play a more important role in disk formation compared to other parameters, like turbulence, misalignment, and magnetic field strengths.
In this case, the disk radii could more depend on the magnetic diffusivities of the non-ideal MHD effects, which are related to cosmic-ray ionization rates and grain size distributions in protostellar sources \citep{Padovani14,Zhao16,Zhao18,Dzyurkevich17,Koga19,Kuffmeier20,Tsukamoto20}. 

Nevertheless, turbulence and misalignment could still prompt angular momentum transfer from large to small scales in protostellar sources. 
The observations of a sample of 20 Class 0 protostars have indeed found that the protostellar envelopes on a 1000 au scale tend to have a higher angular momentum if the magnetic fields are misaligned with the rotational axes \citep{Galametz20}. 
On the other hand, our results show no dependence of the disk radii on the misalignment between the magnetic fields and rotation axes in protostellar sources. 
The combination of these results of the envelope and disk scales could suggest that the self-regulation of disk radii by magnetic braking and diffusion becomes efficient in the vicinity of the disks on a 100 au scale, 
as discussed in theoretical studies incorporating the non-ideal MHD effects \citep{Hennebelle16,Zhao20}.

\subsection{Effects of large-scale motions}\label{discuss2}
In hydrodynamics, disk radii are proportional to initial rotational velocities of parental dense cores \citep{Terebey84,Basu98}. 
The analytical calculations of disk evolution considering magnetic braking and ambipolar diffusion show that disk radii can be insensitive to the rotation of parental cores if the disk rotation is fast enough to keep the toroidal magnetic field stationary at the disk edges \citep{Hennebelle16,Zhao20}. 
The weak dependence of disk radii on the initial core rotation has also been demonstrated with numerical simulations \citep{Zhao20b,Hennebelle20}.
Nevertheless, some other non-ideal MHD simulations show that when the initial core rotation is faster, disk radii tend to be larger \citep{Wurster20}, 
and extended rotating spirals with sizes of 100 au tend to form around central disks, which are transitions from infall- to rotation-dominated regions \citep{Zhao16,Zhao18}. 
Similar spiral features have also been observed in young protostellar disks \citep{Tobin16,Lee20}.

Conventionally, the rotation of dense cores can be assessed with large-scale velocity gradients \citep[e.g.,][]{Goodman93,Caselli02}. 
Our results do not show correlations between the disk radii and the magnitudes and directions of the velocity gradients on the core scale of 0.06 pc (Fig.~\ref{vg}). 
This could suggest that the disk radii are self-regulated by the magnetic braking and diffusion, as discussed in Section~\ref{discuss1}, 
or that the measured velocity gradients do not represent large-scale rotational motion. 
As shown in numerical simulations of formation of dense cores in turbulent environments \citep{Dib10,Zhang18,Verliat20}, 
dense cores may not have uniform rotation, 
and thus the angular momentum of dense cores cannot be directly measured from the large-scale velocity gradients.
Complex velocity structures have indeed been observed in several dense cores, suggesting that there is no uniform rotation in these dense cores \citep{Chen19}. 
In addition, if dense cores are filamentary but not axisymmetric, 
other motions, such as infall, can also induce large-scale velocity gradients \citep{Tobin12}.
Besides, our results show that the magnitudes of our measured velocity gradients are correlated with the non-thermal line widths. 
This correlation could support that the observed velocity gradients on the core scale of 0.06 pc around our sample protostars do not represent large-scale rotation, and are likely related to turbulent motion. 
The VLA observations of the NH$_3$ emission in three Class 0 protostars in Perseus at 3$\arcsec$ resolution ($\sim$800--900 au) have shown filamentary structures within a scale of 10\,000 au, and the power-law indices of the radial profiles of the specific angular momentum are measured to be 1.8, which is between the expected values for rigid-body rotation of 2 and pure turbulence of 1.5 \citep{Pineda19}.
This VLA result demonstrates the need of high-resolution observations to resolve the velocity gradient for proper estimation of the angular momentum in dense cores.
Consequently, a correlation between the disk radii and the observed velocity gradients on the core scale is not expected in the present study, 
and the disk radii could be still related to the net angular momenta of the local velocity fields in individual protostellar sources, as seen in some theoretical simulations \citep{Zhao16,Zhao18,Wurster20}.
%However, the net angular momenta of the dense cores associated with our sample protostars cannot be probed with the data in the present paper.

\subsection{Effects of crowded environments}
Disks formed in a stellar cluster can be truncated or disrupted by dynamical interactions with other stars, 
and disks may also accrete new material after close encounters \citep{Bate12,Bate18, Wurster19}. 
In the hydrodynamics simulations of a turbulent molecular cloud with a mass of 500 $M_\sun$ and a radius of 0.4 pc, 
diverse disks with radii ranging from a few tens to a few hundred au form after 0.2 Myr \citep{Bate18}, 
and the disk radius distribution in the simulations is similar to that in the Orion regions observed with the VANDAM survey \citep{Tobin20}.
In the simulations, more than half of the disks experience dynamical interaction with other stars, 
and the diversity of the disk sizes is due to the initial turbulent environment and the dynamical interactions \citep{Bate12,Bate18}. 
In the MHD simulations of turbulent molecular clouds with a mass of 50 $M_\sun$ and a radius of 0.2 pc, 
diverse disk with radii of ten to a few hundred au also form \citep{Wurster19}. 
In addition, in these simulations with the different initial conditions, such as initial magnetic field strengths and with and without the non-ideal MHD effects, 
the disk radii are all comparable \citep{Wurster19}.
These simulation results could suggest the importance of the dynamical interaction with other stars and the local velocity fields in the disk formation and evolution. 

In our study, we have separated the sample protostars with and without nearby companions with separations smaller than 800 au, 
and the effects of the dynamical interaction in multiple systems are likely excluded in our results. 
However, we cannot exclude the possibility of having a fly-by or close encounters in the past. 
Thus, we cannot distinguish the effects of the non-ideal MHD (Section \ref{discuss1}), the local large-scale gas motions (Section \ref{discuss2}), and the dynamical interaction in a stellar cluster (this section) on the observed disk radius distribution in Orion~A. 
Future studies comparing disk radius distributions in star-forming regions with different environments are needed to further differentiate the importance of the different mechanisms in the disk formation and evolution. 

\section{Summary}
We made use of the publicly available science data from the VANDAM, GAS, and SCUPOL surveys, 
and compared the sizes of a sample of protostellar disks measured in the 0.87 mm continuum emission with the physical properties of their ambient environments on the core scale of 0.6 pc in the Orion~A star-forming region. 
The disk radius was defined to be the 2$\sigma$ width of the Gaussian function fitted to the continuum image.
The typical core scale of 0.6 pc in Orion~A was adopted from the mean size of the dense cores identified in the JCMT 850 $\mu$m data of Orion~A.
For each protostar in our sample, we measured the non-thermal line width and velocity gradient on the core scale using the maps of the kinematic temperature, line width, and centroid line-of-sight velocity from the GAS NH$_3$ data, and we also measured the mean orientation and angular dispersion of the magnetic field on the core scale using the Stokes {\it Q} and {\it U} maps from the JCMT SCUPOL data. 
In addition, we estimated the magnetic field strength and mass-to-flux ratio on the core scale with these data.

We found that the radii of the protostellar disks in multiple systems with smaller separations tend to be smaller, when the separations are smaller than 800 au. 
This trend is likely due to the tidal truncation in multiple systems. 
%In multiple systems with separations larger than 800 au, there is no significant correlation between the disk radii and the separations. 
Thus, we excluded the sources in multiple systems with the separations smaller than 800 au from our sample for the subsequent analysis to avoid any bias due to binary interaction, which results in a subsample of 50 individual protostars.
In our sample, the disk radii do not correlate with the bolometric temperatures of their central protostars. 
For a given range of the bolometric temperatures, the difference in the disk radii is an order of magnitude. 
Thus, we expect that the disk radius distribution observed in the 0.87 mm continuum is not dominantly or solely caused by dust evolution in the disks but is related to other physical mechanisms or a combination of dust evolution with other mechanisms.

In our analysis, we did not find any significant correlation between the disk radii and the physical properties of their ambient environments on the core scales, including non-thermal line width, angular dispersion of the magnetic field orientations, magnetic field strength, mass-to-flux ratio, and misalignment between the magnetic field and the rotational axis of the disk, no matter we considered the entire subsample or only the Class 0 protostars in the subsample.
Our results showing no clear dependence of the disk radii on the turbulence, magnetic field strengths, and misalignment on the core scale are more consistent with the expectation from non-ideal MHD simulations of disk formation in collapsing dense cores, 
where the disk radii are self-regulated by magnetic braking and diffusion of the magnetic field due to the non-ideal MHD effects, 
compared to other simulations which only include turbulence and/or a magnetic field misaligned with the rotational axis.
Therefore, our results could hint that the non-ideal MHD effects play a more important role in the disk formation and evolution than the other physical mechanisms. 

In addition, we did not find any significant dependences of the disk radii on the magnitude of the velocity gradient and alignment between the velocity gradient and the disk major axis. 
In contrast, we found that the non-thermal line width and the magnitude of the velocity gradient show a significant correlation with a Pearson correlation coefficients of 0.68.
Although the disk radii are theoretically expected to be proportional to the rotation of dense cores, 
our results do not show any correlation between the disk radii and the velocity gradients on the core scale.
Therefore, our results could suggest that the observed velocity gradients on the core scale do not trace the large-scale rotation and are more likely related to the turbulent motion.

We note that the protostellar disks in our sample are located in a cluster-forming region. 
Dynamical interaction due to a fly-by or close encounters in a stellar cluster can also affect the disk radius distribution. 
However, such effects cannot be discussed with the current data. 
Future studies comparing disk radius distributions in star-forming regions with different environments are needed to further differentiate the importance of the different mechanisms in the disk formation and evolution.
Besides the disk radii, we have also conducted the same analysis for the continuum fluxes of the protostellar disks in our sample, 
and similarly, no significant one-to-one parameter correlations were found. 

\acknowledgements 
{We thank the VANDAM, GAS, and SCUPOL teams for providing their science data to the public community and for their efforts on making these data easily accessible. 
H.-W.Y. acknowledges support from Ministry of Science and Technology (MOST) in Taiwan through the grant  MOST 108-2112-M-001-003-MY2.
P.M.K. acknowledges support through grants MOST 108-2112-M-001-012 and MOST 109-2112-M-001-022.
}

\clearpage

\begin{appendix}

\section{Measurements in this work}\label{alldata}
Table~\ref{alldata} lists all the measurements obtained from the data and analysis described in Section~\ref{analysis}.

\startlongtable
\begin{longrotatetable}
\begin{deluxetable*}{cccccccccccccccc}
\tablecaption{Sample properties}
\centering
\tablehead{Source & RA & Dec & $T_{\rm bol}$ & $R_{\rm disk}$ & $PA_{\rm disk}$ & $\delta V_{\rm nt}$ & $\theta_{B}$ & $\delta\theta_{B}$ & $B_{\rm pos}$ & $\lambda$ & $\Delta\theta_{\rm disk-B}$ & VG & $\theta_{\rm VG}$ &  $\Delta\theta_{\rm disk-VG}$ \\
 & (J2000) & (J2000) & (K) & (au) & ($\arcdeg$) & (km s$^{-1}$)& ($\arcdeg$) & ($\arcdeg$) & (mG) & & ($\arcdeg$) & (km s$^{-1}$ pc$^{-1}$) & ($\arcdeg$) & ($\arcdeg$)}
\startdata
HOPS-96 & 05:35:29.718 & -04:58:48.60 &  36 &  40 & 135 & 0.625$\pm$0.005 & \nodata & \nodata & \nodata & \nodata & \nodata &  2.77$\pm$0.29 & 233$\pm$6 & 81$\pm$6 \\
HOPS-383 & 05:35:29.785 & -04:59:50.41 &  46 &  60 &  50 & 0.730$\pm$0.015 & \nodata & \nodata & \nodata & \nodata & \nodata &  7.04$\pm$0.71 & 336$\pm$5 & 73$\pm$6 \\
HOPS-95 & 05:35:34.205 & -04:59:52.41 &  42 &  37 &  36 & 0.232$\pm$0.008 & \nodata & \nodata & \nodata & \nodata & \nodata &  1.98$\pm$0.39 & 286$\pm$11 & 70$\pm$12 \\
HOPS-93 & 05:35:15.048 & -05:00:08.05 & 107 & 100 & 122 & 0.358$\pm$0.004 &  62$\pm$26 & \nodata & \nodata & \nodata & 30$\pm$26 &  0.66$\pm$0.16 & 193$\pm$16 & 72$\pm$16 \\
HOPS-92-A-A & 05:35:18.337 & -05:00:32.94 & 186 &  37 & 141 & 0.326$\pm$0.001 &  45$\pm$6 &  6$\pm$2 & 1.01$\pm$0.40 & 1.0$\pm$0.4 &  5$\pm$6 &  1.11$\pm$0.06 & 126$\pm$3 & 15$\pm$4 \\
HOPS-92-A-B & 05:35:18.328 & -05:00:33.18 & 186 & $<$18 & \nodata & 0.326$\pm$0.001 &  47$\pm$7 &  6$\pm$2 & 0.91$\pm$0.34 & 1.1$\pm$0.4 & \nodata &  1.11$\pm$0.06 & 126$\pm$3 & \nodata \\
HOPS-92-B & 05:35:18.270 & -05:00:33.92 & 186 &  57 & 174 & 0.326$\pm$0.001 &  47$\pm$7 &  6$\pm$2 & 0.91$\pm$0.34 & 1.1$\pm$0.4 & 37$\pm$7 &  1.11$\pm$0.06 & 126$\pm$3 & 48$\pm$4 \\
HOPS-91 & 05:35:18.925 & -05:00:51.11 &  42 &  87 &  54 & 0.420$\pm$0.002 &  50$\pm$8 &  6$\pm$2 & 1.32$\pm$0.50 & 0.7$\pm$0.3 & 85$\pm$8 &  3.38$\pm$0.10 &  66$\pm$1 & 12$\pm$2 \\
HOPS-89 & 05:35:19.975 & -05:01:02.56 & 158 &  40 &  82 & 0.390$\pm$0.003 &  40$\pm$7 &  9$\pm$3 & 0.79$\pm$0.31 & 1.3$\pm$0.5 & 48$\pm$7 &  1.44$\pm$0.11 &  61$\pm$4 & 21$\pm$5 \\
HOPS-88 & 05:35:22.471 & -05:01:14.31 &  42 &  33 & 167 & 0.299$\pm$0.003 &  53$\pm$5 &  6$\pm$2 & 0.90$\pm$0.34 & 1.0$\pm$0.4 & 24$\pm$5 &  1.53$\pm$0.14 & 264$\pm$5 & 82$\pm$5 \\
HOPS-87 & 05:35:23.420 & -05:01:30.54 &  38 & 120 &  11 & 0.302$\pm$0.004 &  58$\pm$9 &  8$\pm$2 & 0.74$\pm$0.28 & 1.5$\pm$0.6 & 43$\pm$9 &  1.11$\pm$0.18 & 146$\pm$11 & 44$\pm$12 \\
HOPS-86-A & 05:35:23.654 & -05:01:40.26 & 113 &  57 &   1 & 0.293$\pm$0.004 &  61$\pm$9 &  9$\pm$4 & 0.58$\pm$0.27 & 1.8$\pm$0.8 & 29$\pm$9 &  0.70$\pm$0.21 &  11$\pm$16 & 11$\pm$17 \\
HOPS-86-B & 05:35:23.733 & -05:01:48.13 & 113 &  43 &   7 & 0.286$\pm$0.004 &  49$\pm$12 & \nodata & \nodata & \nodata & 48$\pm$12 &  1.48$\pm$0.18 & 328$\pm$8 & 38$\pm$9 \\
HOPS-85-A & 05:35:28.189 & -05:03:41.20 & 174 &  17 & 124 & 0.338$\pm$0.013 &  65$\pm$23 & 24$\pm$9 & 0.13$\pm$0.06 & 2.0$\pm$0.9 & 31$\pm$23 &  2.29$\pm$0.44 & 332$\pm$14 & 28$\pm$14 \\
HOPS-85-B & 05:35:28.193 & -05:03:41.29 & 174 &  10 &  66 & 0.338$\pm$0.013 &  65$\pm$23 & 24$\pm$9 & 0.13$\pm$0.06 & 2.0$\pm$0.9 & 89$\pm$23 &  2.29$\pm$0.44 & 332$\pm$14 & 86$\pm$14 \\
HOPS-84-A & 05:35:26.560 & -05:03:55.12 &  91 &  87 & 168 & 0.492$\pm$0.005 &  62$\pm$7 &  9$\pm$3 & 0.72$\pm$0.33 & 0.7$\pm$0.3 & 16$\pm$7 &  6.39$\pm$0.26 &  63$\pm$2 & 76$\pm$3 \\
HOPS-84-B & 05:35:26.537 & -05:03:55.73 &  91 &  30 & 166 & 0.492$\pm$0.005 &  62$\pm$7 &  9$\pm$3 & 0.72$\pm$0.33 & 0.7$\pm$0.3 & 14$\pm$7 &  6.39$\pm$0.26 &  63$\pm$2 & 78$\pm$3 \\
HOPS-82 & 05:35:19.740 & -05:04:54.50 & 116 &  20 &  91 & 0.915$\pm$0.016 & \nodata & \nodata & \nodata & \nodata & \nodata & 16.85$\pm$0.68 &  84$\pm$1 &  7$\pm$2 \\
HOPS-81 & 05:35:28.019 & -05:04:57.35 &  40 &  33 & 122 & 0.186$\pm$0.001 &  15$\pm$14 & 12$\pm$4 & 0.15$\pm$0.06 & 1.8$\pm$0.8 & 17$\pm$14 &  2.33$\pm$0.05 & 200$\pm$1 & 78$\pm$1 \\
HOPS-80 & 05:35:25.183 & -05:05:09.37 & 275 &  90 & 155 & 0.201$\pm$0.001 &  18$\pm$14 & 14$\pm$5 & 0.13$\pm$0.06 & 1.9$\pm$0.9 & 47$\pm$14 &  1.35$\pm$0.06 & 113$\pm$2 & 42$\pm$2 \\
HOPS-78-A & 05:35:25.966 & -05:05:43.33 &  38 & 147 & 171 & 0.206$\pm$0.001 &  30$\pm$10 & \nodata & \nodata & \nodata & 52$\pm$10 &  1.14$\pm$0.06 &  21$\pm$2 & 30$\pm$3 \\
HOPS-78-B & 05:35:26.148 & -05:05:45.80 &  38 &  17 & 179 & 0.205$\pm$0.001 &  25$\pm$12 & \nodata & \nodata & \nodata & 65$\pm$12 &  0.99$\pm$0.05 & 339$\pm$2 & 20$\pm$3 \\
HOPS-77-B & 05:35:31.470 & -05:05:46.10 & 550 & $<$18 & \nodata & 0.369$\pm$0.013 &  16$\pm$25 & 18$\pm$7 & 0.17$\pm$0.08 & 1.2$\pm$0.5 & \nodata &  3.00$\pm$0.37 & 100$\pm$9 & \nodata \\
HOPS-78-C & 05:35:26.181 & -05:05:47.12 &  38 &  57 & 132 & 0.205$\pm$0.001 &  25$\pm$12 & \nodata & \nodata & \nodata & 17$\pm$12 &  0.99$\pm$0.05 & 339$\pm$2 & 27$\pm$3 \\
HOPS-77-A-A & 05:35:31.539 & -05:05:47.33 & 550 & $<$18 & \nodata & 0.341$\pm$0.006 &  16$\pm$25 & 18$\pm$7 & 0.17$\pm$0.08 & 1.4$\pm$0.6 & \nodata &  3.26$\pm$0.27 &  17$\pm$4 & \nodata \\
HOPS-77-A-B & 05:35:31.545 & -05:05:47.40 & 550 & $<$18 & \nodata & 0.341$\pm$0.006 &  16$\pm$25 & 18$\pm$7 & 0.17$\pm$0.08 & 1.4$\pm$0.6 & \nodata &  3.26$\pm$0.27 &  17$\pm$4 & \nodata \\
HOPS-78-D & 05:35:25.920 & -05:05:47.70 &  38 &  50 & 175 & 0.238$\pm$0.002 &  18$\pm$6 & \nodata & \nodata & \nodata & 67$\pm$6 &  0.67$\pm$0.07 & 120$\pm$5 & 55$\pm$6 \\
HOPS-76 & 05:35:25.763 & -05:05:58.13 & 136 &  17 &  58 & 0.282$\pm$0.002 &  13$\pm$13 & 11$\pm$4 & 0.31$\pm$0.13 & 1.4$\pm$0.6 & 46$\pm$13 &  2.06$\pm$0.09 & 145$\pm$2 & 88$\pm$3 \\
HOPS-75-A & 05:35:26.680 & -05:06:10.51 &  68 &  20 & 156 & 0.304$\pm$0.002 &  19$\pm$11 &  7$\pm$2 & 0.46$\pm$0.18 & 0.8$\pm$0.3 & 47$\pm$11 &  3.26$\pm$0.11 & 136$\pm$1 & 19$\pm$2 \\
HOPS-75-B & 05:35:26.678 & -05:06:10.76 &  68 & $<$18 & \nodata & 0.304$\pm$0.002 &  19$\pm$11 &  7$\pm$2 & 0.46$\pm$0.18 & 0.8$\pm$0.3 & \nodata &  3.26$\pm$0.11 & 136$\pm$1 & \nodata \\
HOPS-74 & 05:35:24.874 & -05:06:21.58 & 517 &  40 & 127 & 0.392$\pm$0.005 &   5$\pm$17 & 17$\pm$5 & 0.17$\pm$0.07 & 1.0$\pm$0.4 & 32$\pm$17 &  8.02$\pm$0.23 &  79$\pm$1 & 47$\pm$2 \\
HOPS-73 & 05:35:27.699 & -05:07:04.32 &  43 &  63 &  61 & 0.235$\pm$0.005 &  15$\pm$5 & \nodata & \nodata & \nodata & 44$\pm$5 &  0.58$\pm$0.20 &   8$\pm$21 & 52$\pm$22 \\
HOPS-394-A & 05:35:24.046 & -05:07:52.11 &  46 &  83 &  81 & 0.341$\pm$0.002 &  86$\pm$21 & \nodata & \nodata & \nodata & 85$\pm$21 &  1.95$\pm$0.08 & 293$\pm$2 & 33$\pm$3 \\
HOPS-394-B & 05:35:24.867 & -05:07:54.64 &  46 & $<$18 & \nodata & 0.330$\pm$0.002 &  97$\pm$17 & \nodata & \nodata & \nodata & \nodata &  2.73$\pm$0.11 & 281$\pm$2 & \nodata \\
HOPS-71-B & 05:35:25.541 & -05:07:56.84 & 278 &  97 &  82 & 0.343$\pm$0.003 &  97$\pm$17 & \nodata & \nodata & \nodata & 75$\pm$17 &  2.91$\pm$0.11 & 287$\pm$2 & 26$\pm$2 \\
HOPS-71-A & 05:35:25.582 & -05:07:57.64 & 278 &  63 & 123 & 0.343$\pm$0.003 &  97$\pm$17 & \nodata & \nodata & \nodata & 64$\pm$17 &  2.91$\pm$0.11 & 287$\pm$2 & 15$\pm$2 \\
HOPS-70-B-B & 05:35:22.570 & -05:08:00.32 & 619 &  23 &  51 & 0.504$\pm$0.005 &  36$\pm$48 & \nodata & \nodata & \nodata & 74$\pm$48 & 12.09$\pm$0.16 &  51$\pm$0 &  0$\pm$1 \\
HOPS-70-C & 05:35:21.938 & -05:08:04.58 & 619 &  57 &  78 & 0.553$\pm$0.007 &  79$\pm$14 & \nodata & \nodata & \nodata & 90$\pm$14 & 17.39$\pm$0.20 &  45$\pm$1 & 32$\pm$1 \\
HOPS-70-A-B & 05:35:22.426 & -05:08:05.03 & 619 & $<$18 & \nodata & 0.553$\pm$0.007 &  78$\pm$20 & \nodata & \nodata & \nodata & \nodata & 17.39$\pm$0.20 &  45$\pm$1 & \nodata \\
HOPS-70-A-A & 05:35:22.406 & -05:08:05.14 & 619 &  17 &  13 & 0.553$\pm$0.007 &  78$\pm$20 & \nodata & \nodata & \nodata & 25$\pm$20 & 17.39$\pm$0.20 &  45$\pm$1 & 32$\pm$1 \\
HOPS-68 & 05:35:24.299 & -05:08:30.74 & 101 &  33 &   4 & 0.373$\pm$0.002 &  93$\pm$10 & \nodata & \nodata & \nodata &  1$\pm$10 &  0.18$\pm$0.09 & 220$\pm$32 & 37$\pm$33 \\
HOPS-66-B & 05:35:26.927 & -05:09:22.43 & 265 &  57 & 166 & 0.465$\pm$0.005 &  74$\pm$12 & 14$\pm$5 & 0.43$\pm$0.18 & 1.1$\pm$0.5 &  2$\pm$12 &  6.14$\pm$0.22 & 307$\pm$2 & 38$\pm$2 \\
HOPS-66-A & 05:35:26.857 & -05:09:24.40 & 265 &  17 &  47 & 0.465$\pm$0.005 &  74$\pm$12 & 14$\pm$5 & 0.47$\pm$0.20 & 1.3$\pm$0.5 & 63$\pm$12 &  6.14$\pm$0.22 & 307$\pm$2 & 81$\pm$2 \\
HOPS-370 & 05:35:27.634 & -05:09:34.42 &  72 & 113 & 110 & 0.476$\pm$0.006 &  64$\pm$22 & 13$\pm$5 & 0.54$\pm$0.25 & 1.3$\pm$0.6 & 45$\pm$22 &  3.05$\pm$0.22 & 306$\pm$4 & 17$\pm$5 \\
HOPS-65 & 05:35:21.576 & -05:09:38.85 & 546 & 330 & 171 & 0.318$\pm$0.019 & 115$\pm$7 & \nodata & \nodata & \nodata & 34$\pm$7 &  3.55$\pm$0.47 & 142$\pm$9 & 29$\pm$9 \\
HOPS-64 & 05:35:26.998 & -05:09:54.08 &  30 &  33 & 119 & 0.446$\pm$0.002 &  29$\pm$14 & 15$\pm$5 & 0.55$\pm$0.22 & 2.0$\pm$0.8 &  1$\pm$14 &  1.94$\pm$0.10 & 316$\pm$3 & 18$\pm$3 \\
HOPS-108 & 05:35:27.086 & -05:10:00.06 &  39 &  40 & 106 & 0.446$\pm$0.002 &  29$\pm$14 & 15$\pm$5 & 0.55$\pm$0.22 & 2.0$\pm$0.8 & 13$\pm$14 &  1.94$\pm$0.10 & 316$\pm$3 & 31$\pm$3 \\
HOPS-369 & 05:35:26.969 & -05:10:17.27 & 379 &  17 &  21 & 0.433$\pm$0.002 &  19$\pm$8 &  8$\pm$3 & 0.90$\pm$0.38 & 1.0$\pm$0.4 & 88$\pm$8 &  6.67$\pm$0.09 &  13$\pm$0 &  8$\pm$1 \\
HOPS-368 & 05:35:24.725 & -05:10:30.08 & 138 &  63 & 105 & 0.483$\pm$0.005 &  28$\pm$17 & 15$\pm$5 & 0.35$\pm$0.15 & 1.0$\pm$0.4 & 13$\pm$17 & 14.52$\pm$0.13 & 322$\pm$0 & 37$\pm$1 \\
HOPS-60 & 05:35:23.287 & -05:12:03.41 &  54 &  63 & 159 & 0.254$\pm$0.001 &  29$\pm$4 & \nodata & \nodata & \nodata & 40$\pm$4 &  2.65$\pm$0.05 & 323$\pm$1 & 16$\pm$1 \\
HOPS-59-B & 05:35:20.134 & -05:13:13.31 & 528 & $<$18 & \nodata & 0.354$\pm$0.002 & 117$\pm$24 & \nodata & \nodata & \nodata & \nodata &  3.68$\pm$0.09 &  90$\pm$1 & \nodata \\
HOPS-59-A & 05:35:20.151 & -05:13:15.70 & 528 &  83 &  92 & 0.354$\pm$0.002 & 117$\pm$24 & \nodata & \nodata & \nodata & 65$\pm$24 &  3.68$\pm$0.09 &  90$\pm$1 &  1$\pm$2 \\
HOPS-409 & 05:35:21.363 & -05:13:17.83 &  28 & 113 & 116 & 0.417$\pm$0.002 & 137$\pm$29 & \nodata & \nodata & \nodata & 68$\pm$29 &  0.62$\pm$0.12 & 153$\pm$9 & 37$\pm$9 \\
HOPS-58 & 05:35:18.523 & -05:13:38.34 & 620 &  53 & 172 & 0.222$\pm$0.002 & 129$\pm$13 & 13$\pm$4 & 0.12$\pm$0.05 & 1.3$\pm$0.5 & 46$\pm$13 &  0.64$\pm$0.12 & 308$\pm$ 9 & 44$\pm$10 \\
HOPS-57-A & 05:35:19.887 & -05:15:08.18 & 421 &  77 &  43 & 0.265$\pm$0.002 &  45$\pm$8 & 11$\pm$3 & 0.24$\pm$0.10 & 1.3$\pm$0.5 & 88$\pm$8 &  2.17$\pm$0.09 & 357$\pm$2 & 45$\pm$2 \\
HOPS-57-B & 05:35:19.833 & -05:15:09.03 & 421 &  20 &  74 & 0.265$\pm$0.002 &  45$\pm$8 & 11$\pm$3 & 0.24$\pm$0.10 & 1.3$\pm$0.5 & 61$\pm$8 &  2.17$\pm$0.09 & 357$\pm$2 & 76$\pm$2 \\
HOPS-56-A-C & 05:35:19.481 & -05:15:32.81 &  48 & $<$18 & \nodata & 0.330$\pm$0.002 &  35$\pm$9 & 12$\pm$5 & 0.30$\pm$0.14 & 1.2$\pm$0.6 & \nodata &  1.70$\pm$0.12 & 323$\pm$3 & \nodata \\
HOPS-56-A-B & 05:35:19.470 & -05:15:32.96 &  48 & $<$18 & \nodata & 0.330$\pm$0.002 &  33$\pm$11 & \nodata & \nodata & \nodata & \nodata &  1.70$\pm$0.12 & 323$\pm$3 & \nodata \\
HOPS-56-A-A & 05:35:19.482 & -05:15:33.08 &  48 &  23 &  43 & 0.330$\pm$0.002 &  27$\pm$15 & 14$\pm$5 & 0.26$\pm$0.12 & 1.3$\pm$0.6 & 74$\pm$15 &  1.70$\pm$0.12 & 323$\pm$3 & 80$\pm$4 \\
HOPS-56-B & 05:35:19.420 & -05:15:38.29 &  48 & 264 &  61 & 0.330$\pm$0.002 &  26$\pm$12 & 16$\pm$6 & 0.22$\pm$0.10 & 1.5$\pm$0.7 & 55$\pm$12 &  1.70$\pm$0.12 & 323$\pm$3 & 82$\pm$4 \\
HOPS-45-A & 05:35:06.457 & -05:33:34.98 & 518 &  27 &  29 & 1.490$\pm$0.040 & \nodata & \nodata & \nodata & \nodata & \nodata &  6.56$\pm$2.15 & 340$\pm$19 & 49$\pm$19 \\
HOPS-45-B & 05:35:06.454 & -05:33:35.28 & 518 &  17 & 147 & 1.490$\pm$0.040 & \nodata & \nodata & \nodata & \nodata & \nodata &  6.56$\pm$2.15 & 340$\pm$19 & 13$\pm$19 \\
HOPS-44 & 05:35:10.591 & -05:35:06.26 &  44 &  50 & 102 & 0.270$\pm$0.004 & \nodata & \nodata & \nodata & \nodata & \nodata &  3.43$\pm$0.15 &  43$\pm$3 & 58$\pm$4 \\
HOPS-43 & 05:35:04.503 & -05:35:14.78 &  75 &  57 & 120 & 0.175$\pm$0.003 & \nodata & \nodata & \nodata & \nodata & \nodata &  0.78$\pm$0.10 & 149$\pm$6 & 30$\pm$7 \\
HOPS-42 & 05:35:05.061 & -05:35:40.79 & 201 & 133 & 106 & 0.486$\pm$0.010 & \nodata & \nodata & \nodata & \nodata & \nodata & 10.88$\pm$0.30 & 136$\pm$2 & 30$\pm$2 \\
HOPS-40 & 05:35:08.440 & -05:35:58.57 &  38 &  47 &  76 & 0.320$\pm$0.003 & \nodata & \nodata & \nodata & \nodata & \nodata &  2.06$\pm$0.14 &   8$\pm$4 & 67$\pm$4 \\
HOPS-30 & 05:34:44.055 & -05:41:25.84 &  81 &  43 & 140 & 0.294$\pm$0.010 & \nodata & \nodata & \nodata & \nodata & \nodata &  2.52$\pm$0.51 & 188$\pm$11 & 49$\pm$11 \\
HOPS-28-A & 05:34:47.320 & -05:41:56.08 &  46 &  53 &  54 & 0.719$\pm$0.019 & \nodata & \nodata & \nodata & \nodata & \nodata &  3.25$\pm$0.94 & 269$\pm$16 & 35$\pm$17 \\
HOPS-28-B & 05:34:47.303 & -05:41:56.29 &  46 &  27 &  67 & 0.719$\pm$0.019 & \nodata & \nodata & \nodata & \nodata & \nodata &  3.25$\pm$0.94 & 269$\pm$16 & 22$\pm$17 \\
HOPS-17 & 05:35:07.185 & -05:52:06.07 & 341 &  56 &  37 & 0.183$\pm$0.011 & \nodata & \nodata & \nodata & \nodata & \nodata &  5.65$\pm$0.24 & 139$\pm$3 & 77$\pm$3 \\
HOPS-16 & 05:35:00.822 & -05:55:26.07 & 361 &  17 & 148 & 0.270$\pm$0.005 & \nodata & \nodata & \nodata & \nodata & \nodata &  2.38$\pm$0.18 & 111$\pm$3 & 36$\pm$4 \\
HOPS-371 & 05:35:10.419 & -05:55:40.55 &  32 &  43 & 113 & 0.254$\pm$0.002 & \nodata & \nodata & \nodata & \nodata & \nodata &  3.37$\pm$0.09 &  66$\pm$1 & 46$\pm$2 \\
HOPS-12-A & 05:35:08.631 & -05:55:54.64 &  42 &  20 & 103 & 0.486$\pm$0.006 & \nodata & \nodata & \nodata & \nodata & \nodata &  5.49$\pm$0.23 & 183$\pm$2 & 80$\pm$2 \\
HOPS-12-B-B & 05:35:08.955 & -05:55:54.88 &  42 &  30 &  83 & 0.486$\pm$0.006 & \nodata & \nodata & \nodata & \nodata & \nodata &  5.49$\pm$0.23 & 183$\pm$2 & 79$\pm$2 \\
HOPS-12-B-A & 05:35:08.945 & -05:55:55.03 &  42 &  30 &  77 & 0.486$\pm$0.006 & \nodata & \nodata & \nodata & \nodata & \nodata &  5.49$\pm$0.23 & 183$\pm$2 & 74$\pm$2 \\
HOPS-11 & 05:35:13.424 & -05:57:57.88 &  49 &  43 &  94 & 0.272$\pm$0.004 & \nodata & \nodata & \nodata & \nodata & \nodata &  3.03$\pm$0.18 & 295$\pm$3 & 21$\pm$4 \\
HOPS-10 & 05:35:09.050 & -05:58:26.87 &  46 &  46 & 117 & 0.179$\pm$0.002 & \nodata & \nodata & \nodata & \nodata & \nodata &  1.64$\pm$0.11 & 282$\pm$4 & 14$\pm$4 \\
HOPS-200 & 05:35:33.214 & -06:06:09.62 & 244 & 151 &  84 & 0.903$\pm$0.037 & \nodata & \nodata & \nodata & \nodata & \nodata & 11.45$\pm$1.93 & 245$\pm$10 & 18$\pm$10 \\
HOPS-194 & 05:35:52.027 & -06:10:01.67 & 645 &  62 &  87 & 0.537$\pm$0.017 & \nodata & \nodata & \nodata & \nodata & \nodata &  2.05$\pm$0.79 & 322$\pm$22 & 56$\pm$22 \\
\enddata 
\tablecomments{$T_{\rm bol}$ is the bolometric temperature \citep{Furlan16}. $R_{\rm disk}$ is the disk radius, which is defined as the 2$\sigma$ width of the Gaussian function fitted to the continuum emission \citep{Tobin20}. $PA_{\rm disk}$ is the position angle of the disk major axis after deconvolution, which is defined as the angle increasing from the north to the east. $\delta V_{\rm nt}$ is the non-thermal line width on the core scale of 0.6 pc. $\theta_{B}$ is the mean magnetic field orientation on the core scale. $\delta\theta_{B}$ is the angular dispersion of the magnetic field orientations on the core scale. $B_{\rm pos}$ is the magnetic field strength on the core scale. $\lambda$ is the dimensionless mass-to-flux ratio. $\Delta\theta_{\rm disk-B}$ is the misalignment angle between the magnetic field on the core scale and the rotational axis of the disk, where 90$\arcdeg$ means orthogonal. VG is the magnitude of the velocity gradient on the core scale. $\theta_{\rm VG}$ is the position angle of the direction of the velocity gradient on the core scale. $\Delta\theta_{\rm disk-VG}$ is the alignment angle between the direction of the velocity gradient on the core scale and the disk major axis, where 0$\arcdeg$ means the two are aligned.}
%Reference.
\end{deluxetable*}\label{}
\end{longrotatetable}
\end{appendix}

\end{document}